\documentclass{article}

     \usepackage[final]{neurips_2023}

\usepackage[utf8]{inputenc} 
\usepackage[T1]{fontenc}    
\usepackage{hyperref}       
\usepackage{url}            
\usepackage{booktabs}       
\usepackage{amsfonts}       
\usepackage{nicefrac}       
\usepackage{microtype}      
\usepackage{xcolor}         
\usepackage{graphicx}      
\usepackage{amsmath}       

\title{Precipitation Nowcasting With Spatial And Temporal Transfer Learning Using Swin-UNETR}

\author{%
  Ajitabh Kumar\\
  Enrflo\\
  Noida, UP 201301 India \\
  \texttt{ajitabh.kumar@enrflo.com} \\
}

\begin{document}

\maketitle

\begin{abstract}
  Climate change has led to an increase in frequency of extreme weather events. Early warning systems can prevent disasters and loss of life. Managing such events remain a challenge for both public and private institutions. Precipitation nowcasting can help relevant institutions to better prepare for such events. Numerical weather prediction (NWP) has traditionally been used to make physics based forecasting, and recently deep learning based approaches have been used to reduce turn-around time for nowcasting. In this work, recently proposed Swin-UNETR (Swin UNEt TRansformer) is used for precipitation nowcasting for ten different regions of Europe. Swin-UNETR utilizes a U-shaped network within which a swin transformer-based encoder extracts multi-scale features from multiple input channels of satellite image, while CNN-based decoder makes the prediction. Trained model is capable of nowcasting not only for the regions for which data is available, but can also be used for new regions for which data is not available.
\end{abstract}

\section{Introduction}

Extreme events are more frequent with climate-change, and managing them has become a challenge for various institutions. Precipitation forecasting is required for planning not only normal activities but also mitigating disasters. Traditionally NWP based approaches have been used for precipitation forecasting, but they require relatively large turn around time which makes them ineffective for nowcasting. Deep learning based approaches on the other hand require high compute resources while training, but trained model can yield predictions within seconds using available satellite and radar images [1, 2]. Recently many advances have been made in deep learning based nowcasting and such approaches have even outperformed NWP in nowcasting.

Precipitation nowcasting competition, Weather4Cast 2023, was organized in conjunction with NeurIPS 2023 conference with challenge problem of predicting next 8 hour precipitation prediction over ten different regions of Europe. Satellite images based training data is available for only seven out of ten regions of Europe for the first two years (i.e. 2019 and 2020). Spatial transfer learning is required for nowcasting precipitation for remaining three regions for which no training data is provided. Finally, temporal transfer learning is required for nowcasting precipitation for all ten regions for the third year (i.e. 2021).

Precipitation forecasting and nowcasting is useful for many different activities including agriculture, transport etc., as well as disaster mitigation due to sudden flooding. Nowcasting becomes even more important with increasing frequency of extreme events. But precipitation nowcasting is complex exercise as such events are extremely heterogeneous with varying degree of intensities. Precipitation is based on factors with multiple-scales which makes learning from data quite challenging. We have used recently proposed Swin UNETR architecture to solve multi-scale nowcasting problem including spatio-temporal transfer learning [3-6].

\section{Method}

In this section we first give details of training database used in this study, and then discuss the Swin UNETr architecture used for solving the nowcasting problem.

\subsection{Dataset}

Ground-radar reflectivity measurements are used to calculate pan-European composite rainfall rates by the Operational Program for Exchange of Weather Radar Information (OPERA) radar network. Even though these are more precise, accurate, and of higher resolution than satellite data, they are expensive to obtain and not available in many parts of the world. Hence, goal of this competition is to learn how to predict this high value rain rates from radiation measured by geostationary satellites operated by the European Organisation for the Exploitation of Meteorological Satellites (EUMETSAT).

\begin{figure}
  \centering
  \includegraphics[width=0.8\linewidth]{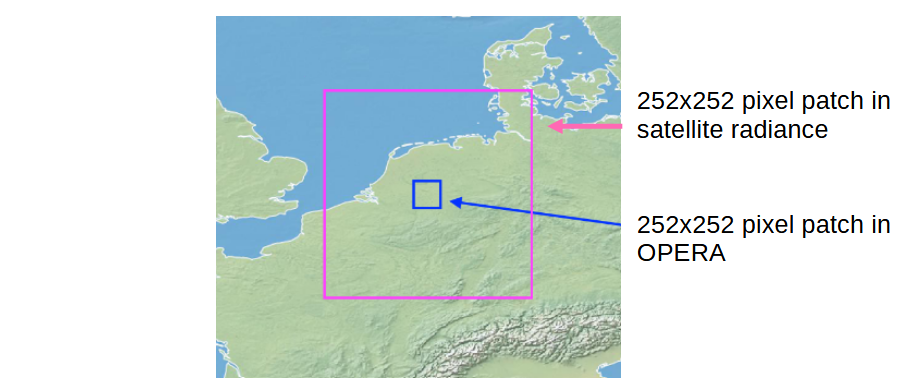}
  \caption{Central area of interest within larger context area for a region under study}
  \label{context}
\end{figure}

Data is available for a total of ten regions of Europe of which seven has both training and test data, while three regions only has test data. Input feature set consists of satellite based images with $11$ different channels for larger context region of area $3024\times3024$ km. Image contains a data point for every grid size of $12\times12$ km, and hence image is of $252\times252$ dimensions. Channels denote slightly noisy satellite radiance covering so-called $2$ visible (VIS), $2$ water vapor (WV), and $7$ infrared (IR) bands. Precipitation data on the other hand is available for the central $504\times504$ km area with a grid size of $2\times2$ km. Thus, even precipitation images are of $252\times252$ dimension. Figure~\ref{context} shows the larger context area with satellite data and the central area with OPERA based precipitation data for one of the regions. Input data is cleaned in the pre-processing step in which all unavailable values are replaced with $0$. Input values are normalized so that different channels could be used together. Deep learning core task is to predict high-resolution ($2\times2$ km) precipitation values using low resolution ($12\times12$ km) satellite images.

\subsection{Swin UNETR Architecture}

Swin UNETR (swin UNEt TRansformers) architecture contains two main modules: a swin transformer-based encoder and a CNN-based decoder [5, 6]. Both are connected in a U-shaped network via skip connections at different resolutions. Attention mechanism in swin transformer is quite effective as it uses hierarchical windows of multiple sizes to handle different scales within feature set. 3D patches at one level is shifted in the next level, thus learning multi-scale contextual representations and long-range dependencies. Encoder has $4$ stages, each with $2$ transformer blocks for regular and shifted partitioning. Extracted feature representations of the encoder are used in the CNN-based decoder via skip connection at each resolution.

In subsequent layers $l$ and $l+1$ of the encoder, outputs are calculated as:

\begin{equation} 
\begin{split} 
\hat{z}^l = WMSA(LN(z^{l-1})) + z^{l-1} \\
z^l = MLP(LN(\hat{z}^l)) + \hat{z}^l \\
\hat{z}^{l+1} = SWMSA(LN(z^{l})) + z^{l} \\
z^{l+1} = MLP(LN(\hat{z}^{l+1})) + \hat{z}^{l+1}
\end{split} 
\end{equation} 

where $WMSA$ and $SWMSA$ are regular and window partitioned multi-head self-attention modules respectively; $\hat{z}^l$ and $\hat{z}^{l+1}$ denote the outputs of $WMSA$ and $SWMSA$; $MLP$ and $LN$ denote layer normalization and multi-layered perceptron respectively.

\section{Results}

Original input data consists of low-resolution satellite image with $11$ different channels for a larger context area, while high-resolution prediction is done for a smaller central area. We use Adam optimizer and cosine annealing learning rate with linear warm-up. Mean squared error is used as loss function in the initial epochs, which is changed to weighted mean absolute error in the later epochs. Pixel-wise weight is calculated as $min(1+y, 24)$ where $y$ is the target precipitation rate. This helps in learning less frequent but high value precipitation events, and ensuring that network does not only fit to dominant medium to low value precipitation events.

\begin{figure}
  \centering
  \includegraphics[width=0.8\linewidth]{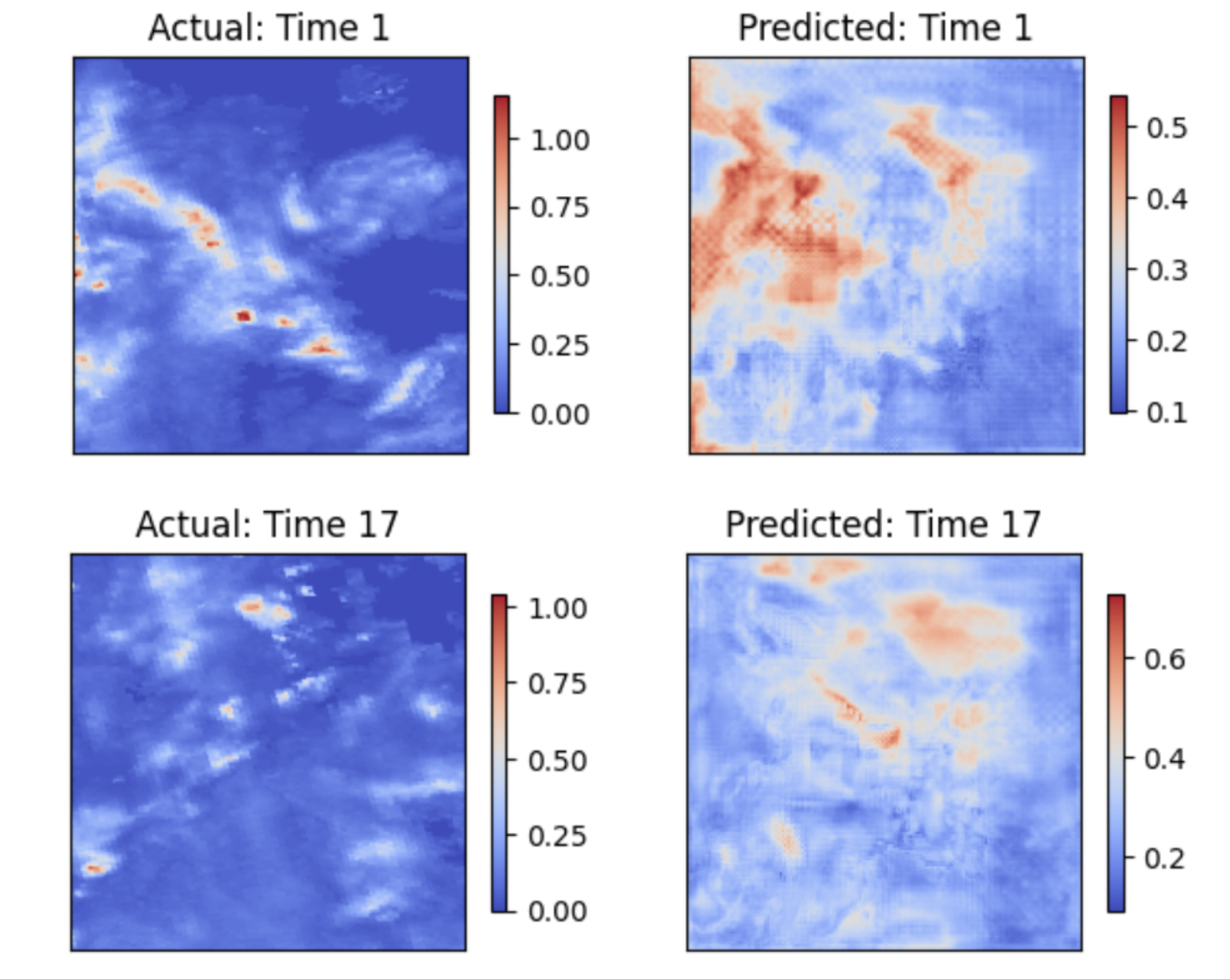}
  \caption{Actual and predicted rainfall values for different prediction times (case 1)}
  \label{case1}
\end{figure}

\begin{figure}
  \centering
  \includegraphics[width=0.8\linewidth]{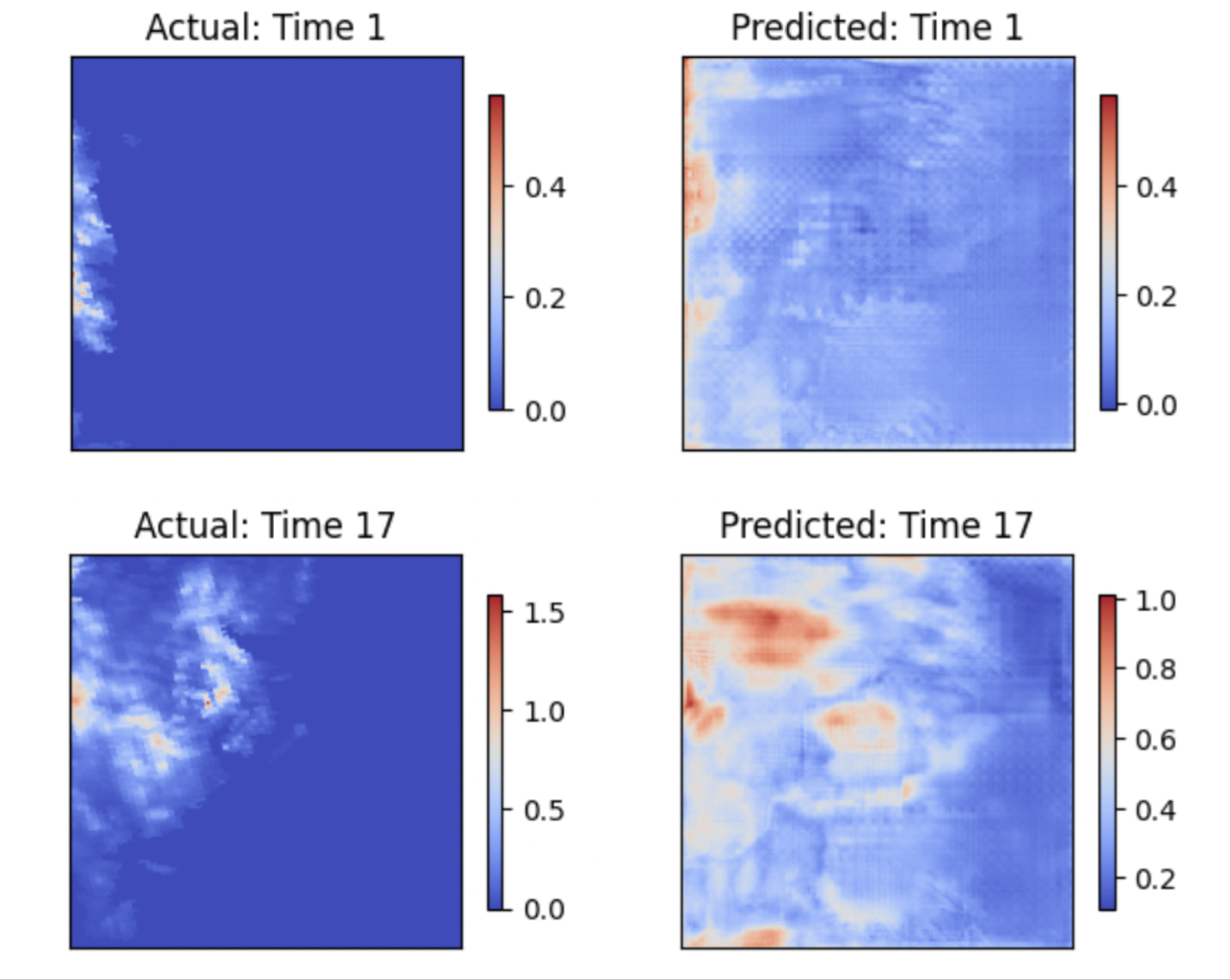}
  \caption{Actual and predicted rainfall values for different prediction times (case 2)}
  \label{case2}
\end{figure}

For the core challenge, input data consists of $4$ past values at $15$-minute interval (i.e. past $1$ hour), while prediction is made for next $32$ future times at $15$-minute interval (i.e. next $8$ hours). Next for the nowcasting and transfer-learning challenge, input data consists of $4$ past values at $15$-minute interval (i.e. past $1$ hour), while prediction is made only for the next $16$ future times at $15$-minute interval (i.e. next $4$ hours). First, we use the previously learnt model and fine-tune it for the $16$ output values. Figures ~\ref{case1} and ~\ref{case2} show the actual and predicted rainfall values for two cases. Model is able to capture the rain events. Due to the size of model, batch size was kept $2$ which made the computation slow. Longer training with more epochs could improve results further. Pre-processing data appropriately to capture logarithmic trend and non-stationarity could also improve results.

\section{Conclusions}

Swin UNETR architecture is used to train neural network model for precipitation nowcasting using multiple channels of satellite images. Swin transformer based encoder learns multi-scale features, while CNN based decoder makes prediction in a U-shaped network. Model outperforms U-Net based baseline and makes good predictions.

\begin{ack}
This research is made possible by generous cloud credits provided by OVH Cloud.
\end{ack}

\section*{References}

\medskip

{
\small

[1] Asperti, A., Merizzi, F., Paparella, A., Pedrazzi, G., Angelinelli, M.\ \& Colamonaco, S.\ (2023) Precipitation nowcasting with generative diffusion models. {\it arXiv preprint arXiv:2308.06733}.

[2] Ravuri, S., Lenc, K., Willson, M. et al.\ (2021) Skilful precipitation nowcasting using deep generative models of radar. {\it Nature} {\bf 597}:672–677.

[3] Liu, Z., Lin, Y., Cao, Y., Hu, H., Wei, Y., Zhang, Z., Lin, S.\ \& Guo, B.\ (2021) Swin transformer: Hierarchical vision transformer using shifted windows. {\it IEEE/CVF International Conference on Computer Vision (ICCV)}.

[4] Hatamizadeh, A., Tang, Y., Nath, V., Yang, D., Myronenko, A., Landman, B., Roth, H.\ \& Xu, D.\ (2021) UNETR: Transformers for 3D Medical Image Segmentation. {\it arXiv preprint arXiv:2103.10504}.

[5] Hatamizadeh, A., Nath, V., Tang, Y., Yang, D., Roth, H.R.\ \& Xu, D. (2022). Swin UNETR: Swin transformers for semantic segmentation of brain tumors in MRI images.  {\it arXiv preprint arXiv:2201.01266}.

[6] Belousov, Y., Polezhaev, S.\ \& Pulfer, B.\ (2022) Solving the Weather4cast challenge via visual transformers for 3D images. {\it arXiv preprint arXiv:2212.02456}.

}


\end{document}